\documentclass[11pt]{article}
\usepackage{moriond,epsfig}

\def\Journal#1#2#3#4{{#1} {\bf #2}, #3 (#4)}

\def\PRL{\em Phys. Rev. Lett.}

\def\PRC{{\em Phys. Rev.} C}

\def\NPA{{Nucl. Phys.}~{\bf A}}

\def\be{\begin{equation}}
\def\ee{\end{equation}}
\def\bea{\begin{eqnarray}}
\def\eea{\end{eqnarray}}

\begin{document}
\vspace*{4cm}
\title{FIRST RESULTS FROM PHENIX AT RHIC}

\author{ A. BAZILEVSKY for the PHENIX Collaboration }

\address{RIKEN BNL Research Center, Brookhaven National Laboratory\\
Upton, NY 11973-5000, USA}

\maketitle\abstracts{
The first results from Au-Au collisions at $\sqrt{s_{NN}}$=130 GeV 
obtained with the PHENIX detector in the Year 2000 run at RHIC are presented. 
The mid-rapidity charged particle multiplicity and transverse energy 
per participating nucleon rise steadily with the number of 
participants, such that $\langle E_T\rangle/\langle N_{ch}\rangle$ remains 
relatively constant as a function of centrality. 
Identified charged hadron spectra as well as $\bar{p}/p$ and $K^+/K^-$ ratios 
are discussed. 
Charged particle and neutral pion transverse momentum distributions in 
peripheral nuclear collisions are consistent with point-like scaling. 
The spectra at high $p_t$ from central collisions are significantly suppressed when compared to a simple superposition of binary nucleon-nucleon collisions. 
}

\section{Introduction}

The Relativistic Heavy-Ion Collider (RHIC) at Brookhaven National
Laboratory started regular operation in June 2000, opening new frontiers
in the study of hadronic matter under unprecedented conditions of
temperature and energy density. 

PHENIX~\cite{ph} is one of the four experiments at RHIC. It is designed to 
simultaneously measure a wide range of probes -- photons, leptons and 
hadrons -- sensitive to all timescales, from initial hard scattering to 
final state interactions. 

All results presented in this paper are obtained using the two PHENIX central 
arm spectrometers covering approximately $\pm0.35$ in pseudorapidity and 
$\pi$ in azimuth. A highly segmented Electromagnetic calorimeter (EMCal) 
allowed measurements of 
photons and neutral pions and also provided global event characterization 
through transverse energy. Charged particles were reconstructed using the 
Drift Chambers (DC), Pad Chambers (PC) and Time-of-Flight (ToF) walls. 
The momentum resolution achieved so far with DC is 
$\sigma_p/p^2=3.5\%/(GeV/c)$ for particles with momentum above 1 GeV/c. 
Particle identification was performed by ToF measurements. 
A large Ring Imaging Cherenkov counter (RICH) was used for electron 
identification. 

Two other detectors, Beam-Beam Counters (BBC) and Zero Degree Calorimeters 
(ZDC), provided a trigger and the off-line event selection~\cite{mult}. 
The BBC comprise two arrays of 64 quartz radiator Cherenkov detectors, located at $\pm$1.44m from the center of the interaction region (IR) covering 2$\pi$ in $\phi$ and $3.0<|\eta|<3.9$. The BBC also provide the start timing and vertex position~\cite{bbc}. The ZDC are small transverse-area hadron calorimeters that measure neutron energy within $|\eta|>6$. 
The centrality bins were determined based upon the BBC vs. ZDC signal distribution. Simulations of the response of the BBC and the ZDC were used to account for the effect of physics and detector fluctuations in the definition of these event classes and to relate them via a Glauber model to the number of participating nucleons $N_{p}$ and the number of binary collisions $N_{c}$ ~\cite{mult}.

\begin{figure}[t]
\vspace{-0.3in}
\begin{minipage}[t]{79mm}
\psfig{figure=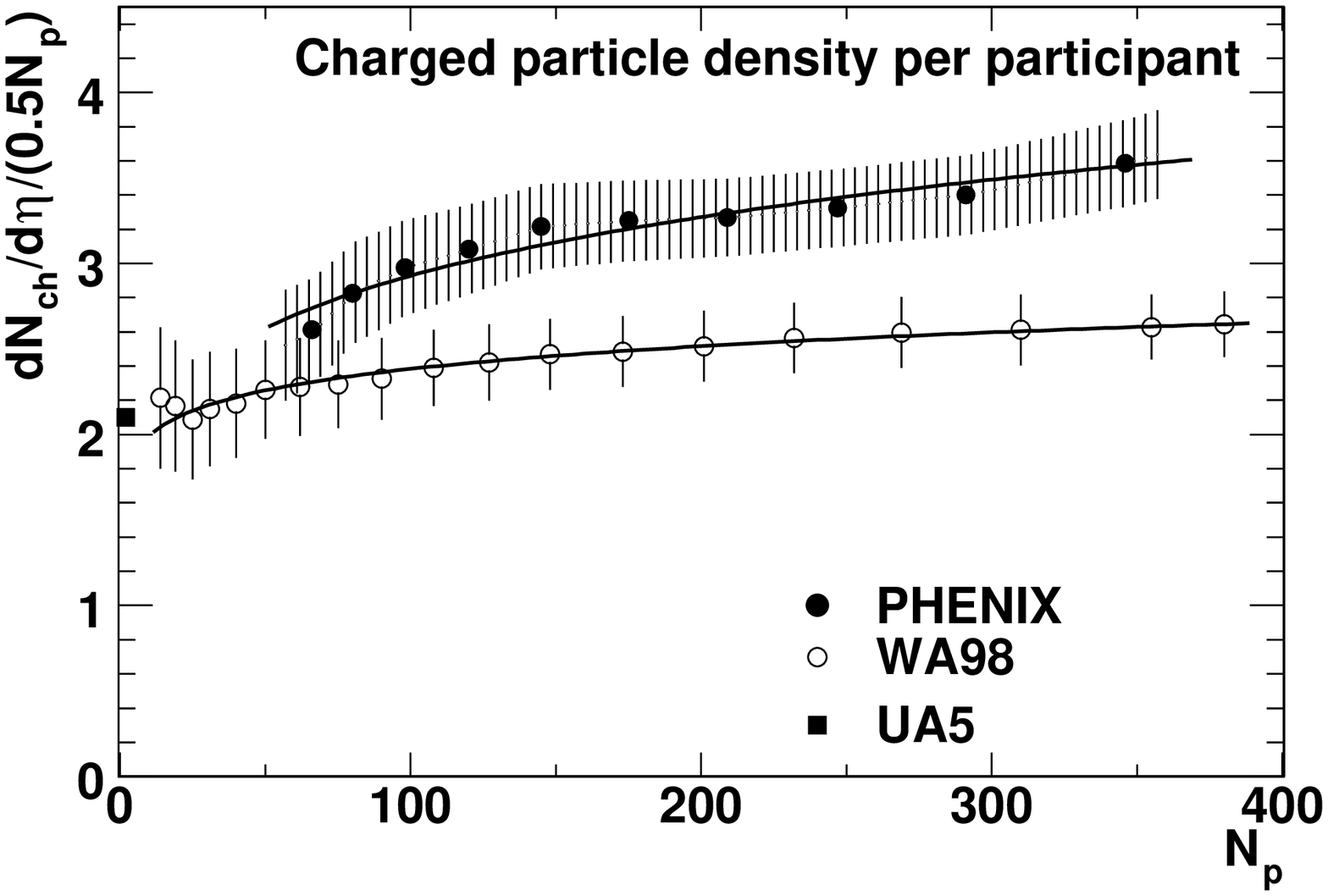,height=2.0in}
\end{minipage}
\begin{minipage}[t]{79mm}
\psfig{figure=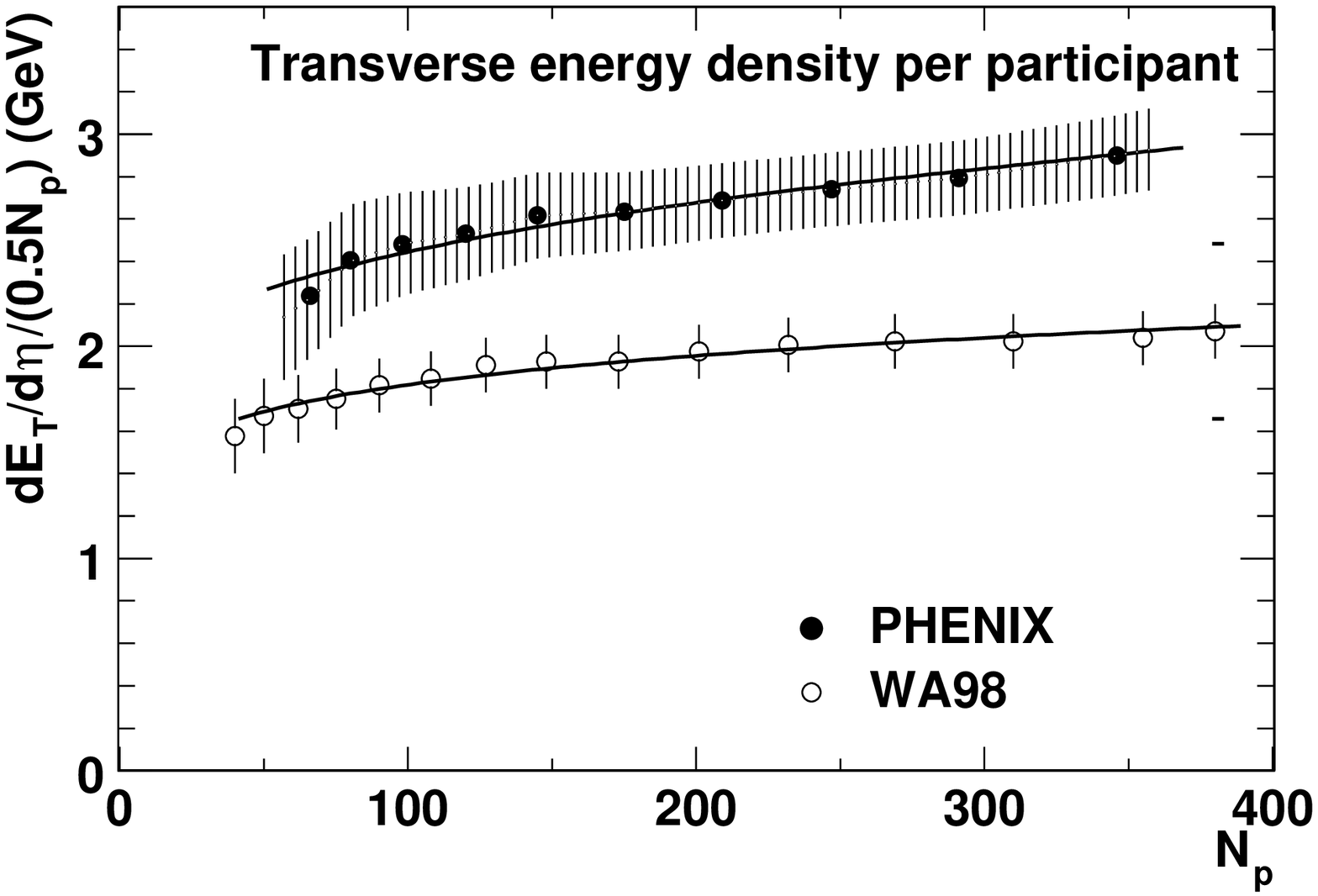,height=2.0in}
\end{minipage}
\caption{Multiplicity (left) and Transverse energy (right) densities per participant pair at mid-rapidity as function of the number of participants.
\label{fig:et1}}
\end{figure}

\section{Charged particle multiplicity and transverse energy}

The number of primary charged particles per event was determined on a statistical basis by correlating two layers of Pad Chambers placed at radial distances of 2.49~m and 4.98~m, respectively, from the IR~\cite{mult}. A section of PbSc Calorimeter  provided the transverse energy measurements~\cite{et}. Fig.1 shows the results for the multiplicity and transverse energy densities per pair of participating nucleons as a function of $N_{p}$. Both quantities exhibit very consistent behavior, such that transverse energy per charged particle, $\langle E_T\rangle/\langle N_{ch}\rangle$, remains relatively constant at a value $\sim 0.8$ GeV (Fig.2 left). The same value for the most central collisions can be derived from the measurements done by WA98~\cite{wa98} and NA49~\cite{na49} at SPS at $\sqrt{s_{NN}}$=17.2 GeV and from the measurement of E814/E877~\cite{e814} at AGS at $\sqrt{s_{NN}}$=4.8 GeV (Fig.2 right). Coming back to Fig.1, it should be noted that the slope parameters $\alpha$ in $dN_{ch}/d\eta$ and $dE_{T}/d\eta$ parameterizations $\propto N_{p}^{\alpha}$ obtained at RHIC~\cite{mult,et,milov} are larger than at SPS: our data give $\alpha=1.16\pm0.04$ ($\alpha=1.13\pm0.05$) for multiplicity density (transverse energy density) compared to $\alpha=1.07\pm0.04$ ($\alpha=1.08\pm0.06$) found by the WA98 experiment~\cite{wa98}. That allowes us to state that at RHIC energy the ``wounded nucleon model'' ($\alpha=1$) breaks down and a term proportional to the number of collisions makes a significant contribution to particle and energy production and it is larger than at SPS where $\alpha$ is found to be close to unity within the errors of measurements. Fig.1 (left) also shows $dN_{ch}/d\eta|_{\eta=0}$ for $N_p=2$ from UA5~\cite{ua5} data for $p\bar{p}$.

\begin{figure}[b]
\begin{minipage}[t]{79mm}
\psfig{figure=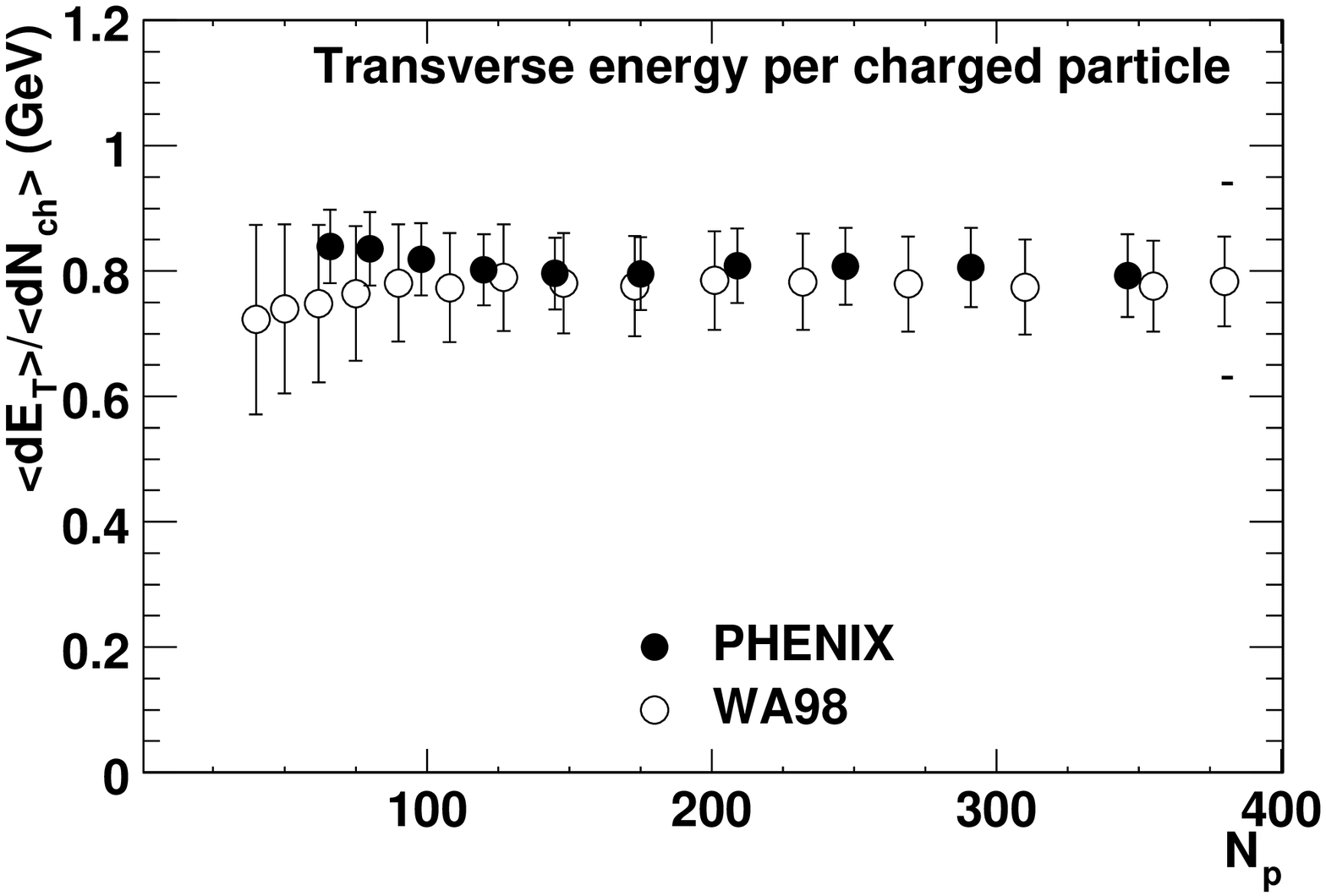,height=2.0in}
\end{minipage}
\begin{minipage}[t]{79mm}
\psfig{figure=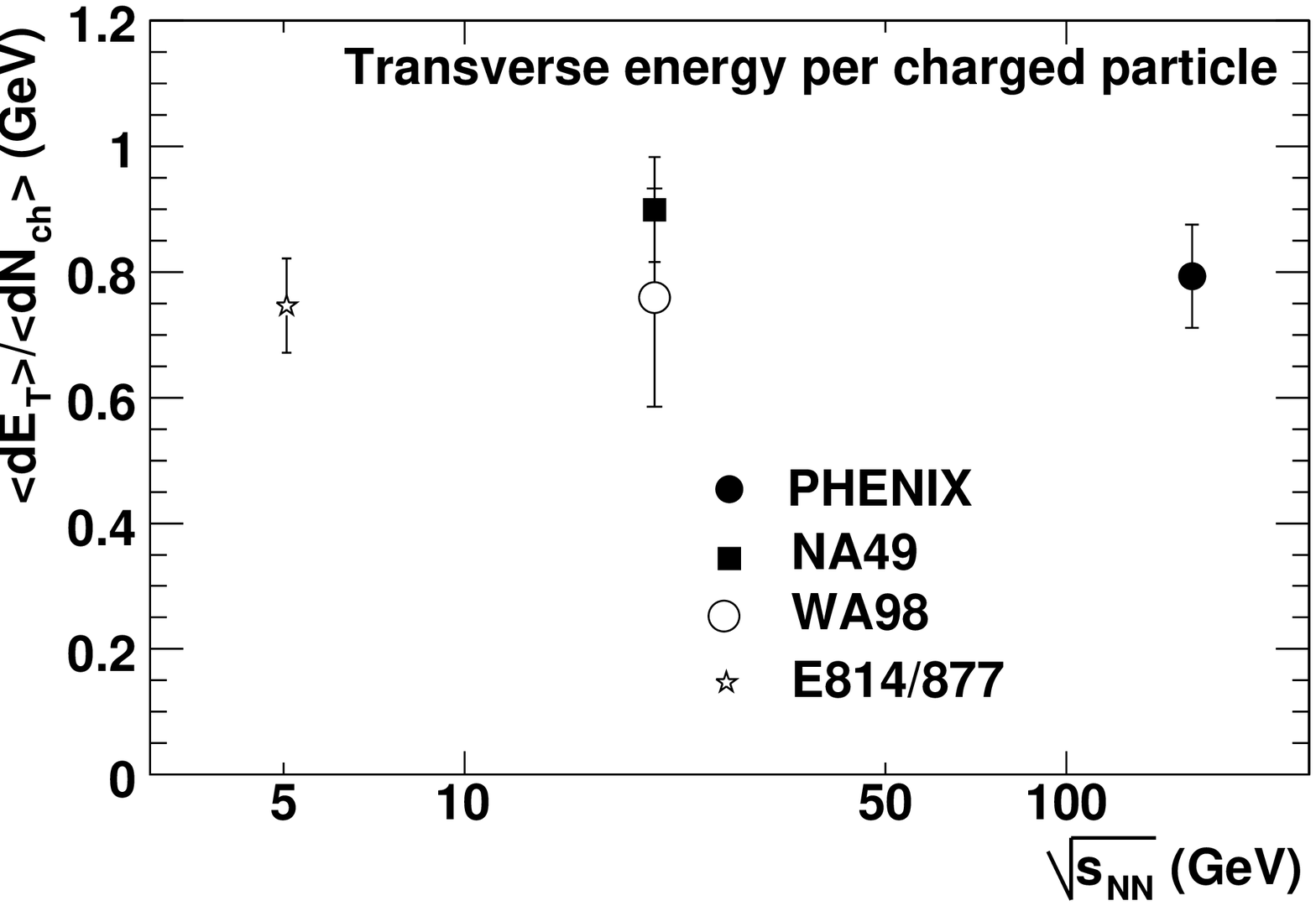,height=2.0in}
\end{minipage}
\caption{Transverse energy per charged particle at mid-rapidity vs $N_{part}$ (left) and energy of participating nucleons (right) for the highest centrality bin.
\label{fig:et2}}
\end{figure}

Our straightforward derivation of Bjorken energy density $\epsilon_{Bj}$ from our measured $dE_T/d\eta|_{\eta=0}$ (and corrected to $dE_T/dy|_{y=0}$) for the most central 2\% of the inelastic cross section gives $\epsilon_{Bj}=4.6$~GeV/fm$^3$ (ref.~\cite{et}), an increase of more than 60\% over the NA49 value~\cite{na49et}. 

\begin{figure}[t]
\vspace{-0.4in}
\begin{center}
\psfig{figure=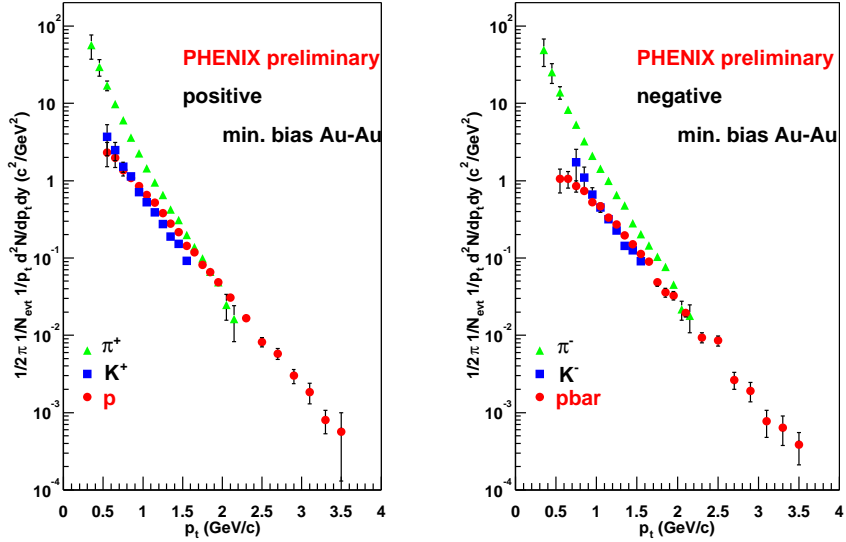,width=4.8in}
\end{center}
\caption{Minimum bias transverse momentum distribution for positive (left) and negative (right) identified hadrons. The error bars include statistical errors and systematic errors in the acceptance and decay corrections. Additional 20\% systematic errors on the absolute normalization are not included.
\label{fig:hadron}}
\end{figure}

\section{Hadron yields}

The ToF time resolution of 115 ps allowed pion and kaon separation 
up to 1.5--2 GeV/c, and proton separation up to 3.5 GeV/c. 
Fig.3 shows the minimum bias $p_t$ spectra for $\pi^{+/-}$, $K^{+/-}$, $p$ and $\bar{p}$. The inverse slope parameters increase with particle mass~\cite{julia}, such that protons and antiprotons become the significant contributors to hadron $p_t$ spectra above 2 GeV/c. 

The $\bar{p}/p$ and $K^+/K^-$ ratios for minimum bias collisions was found to be $0.64\pm0.01(stat)\pm0.07(syst)$ and $1.08\pm0.03(stat)\pm0.22(syst)$ correspondingly. Within uncertainties, neither transverse momentum dependence nor centrality dependence were seen in the $\bar{p}/p$ and $K^+/K^-$ ratios~\cite{hiroaki}. At RHIC energies both $\bar{p}/p$ and $K^+/K^-$ ratios arec learly approaching unity, indicating that pair creation of $\bar{p}/p$ and $K^+/K^-$ in the midrapidity region is becoming a dominant process and implying that the net baryon density is much less than that measured in lower energy experiments at AGS~\cite{ags1,ags2} and SPS~\cite{sps}. 

\section{High transverse momentum particles}

In heavy ion collisions at RHIC energies jet production is expected to dominate particle production at high $p_t$. In the high density medium created in central heavy ion collisions it is predicted that scattered partons may lose considerable energy via gluon bremsstrahlung or ``jet quenching''~\cite{quench1,quench2}. It may result in a reduction of the high $p_t$ hadron yield. 

Fig.4 shows the transverse momentum distribution of $\pi^0$'s in peripheral (upper 60-80\% of $\sigma_{int}$) and the 10\% most central Au-Au collisions. Both spectra are compared to different theoretical calculations~\cite{quench1}. The peripheral data are consistent with pQCD calculations for $pp$, with simple point-like scaling to Au-Au collisions by the mean number of binary collisions. By contrast, the central data points are well below curves calculated without acounting for parton energy loss~\cite{gabor}. The significant suppression of high $p_t$ particles in central collisions was also observed in charged hadron spectra~\cite{federika}. 

\begin{figure}[hbt]
\vspace{-0.4in}
\begin{center}
\psfig{figure=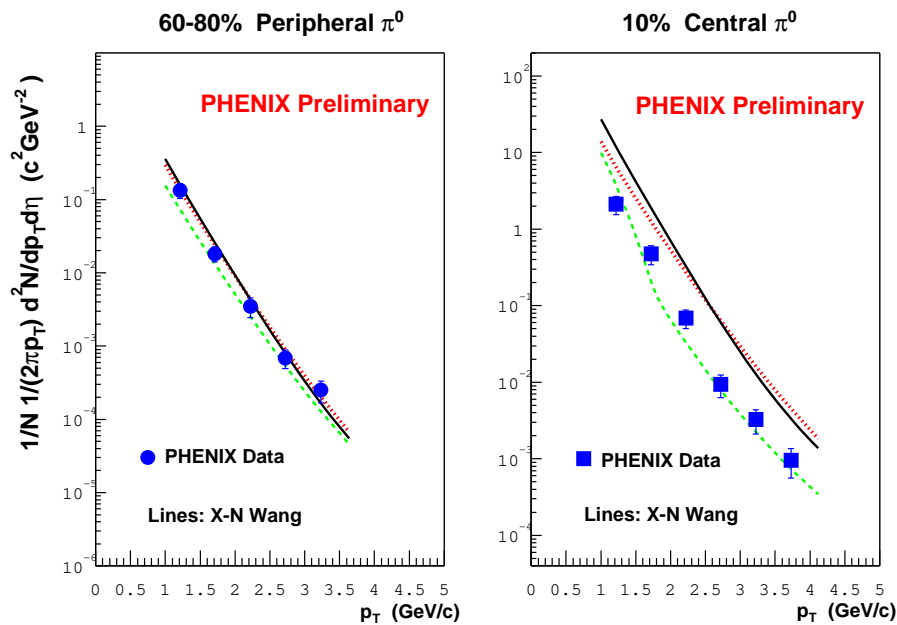,width=4.8in}
\end{center}
\caption{Comparison of $\pi^0$ spectra to theoretical calculations under three scenarios and for two centralities. The points are PHENIX data. The curves are calculations of X-N. Wang: solid lines are a pQCD calculation for $pp$ scaled by the number of binary collisions; the dotted lines add shadowing and $p_t$ broadening; the dashed lines add a $dE/dx=0.25$ GeV/fm parton energy loss.
\label{fig:pi0}}
\end{figure}

\section{Conclusion}

In just a few months after the first Au-Au collisions were observed at RHIC in June 2000, the PHENIX collaboration produced the first intriguing physics results. Because of the limited size of the paper we presented here only some of the PHENIX results regarding charged multiplicity and transverse energy distributions, hadron yields and spectra of high transverse momentum particles. One of the unique PHENIX capabilities is electron identification in a wide momentum range. The first inclusive electron spectra and electron pair mass distribution were obtained from year 2000 data~\cite{akiba}. In the coming run in 2001, PHENIX will accumulate data of about three orders of magnitude larger, so that we expect to observe several thousand $J/\psi,\phi \rightarrow e^+e^-$ events.

\section*{Acknowledgments}

We thank the staff of the RHIC project, Collider-Accelerator, and Physics
Departments at BNL and the staff of PHENIX participating institutions for
their vital contributions.  We acknowledge support from the Department of
Energy and NSF (U.S.A.), Monbu-sho and STA (Japan), RAS, RMAE, and RMS
(Russia), BMBF and DAAD (Germany), FRN, NFR, and the Wallenberg Foundation
(Sweden), MIST and NSERC (Canada), CNPq and FAPESP (Brazil), IN2P3/CNRS
(France), DAE (India), KRF and KOSEF (Korea), and the US-Israel Binational
Science Foundation.

\end{document}